\documentclass[aps,twocolumn,preprintnumbers,amsmath,amssymb]{revtex4}
\input{epsf}

\usepackage{graphicx}
\usepackage{dcolumn}
\usepackage{bm}
\usepackage{graphicx}
\usepackage[normalem]{ulem}
\usepackage{natbib}
\usepackage{color}

\def\lsim{\,{}_\sim^{<}\,}

\begin{document}

\title{Stringent Limit on Primordial Magnetic Fields from the Cosmic Microwave Background Radiation}

\author{
  Karsten Jedamzik$^{\it \bf 1}$\footnote{E-mail: karsten.jedamzik@umontpellier.fr},
  Andrey Saveliev$^{\it \bf 2,3}$\footnote{E-mail: andrey.saveliev@desy.de}
}

\affiliation{
$^{\bf \it 1}$\mbox{\footnotesize Laboratoire Univers et Particules de Montpellier, UMR5299-CNRS, Universit\'e de Montpellier, 34095 Montpellier, France}\\
$^{\bf \it 2}$\mbox{\footnotesize Institute of Physics, Mathematics and Information Technology, Immanuel Kant Baltic Federal University, 236016 Kaliningrad, Russia} \\
$^{\bf \it 3}$\mbox{\footnotesize Faculty of Computational Mathematics and Cybernetics, Lomonosov Moscow State University, 119991 Moscow, Russia}
}

\begin{abstract}
Primordial magnetic fields (PMFs), being present before the epoch of cosmic recombination, induce small-scale baryonic density fluctuations. These inhomogeneities lead to an inhomogeneous recombination process that alters the peaks and heights of the large-scale anisotropies of the cosmic microwave backround (CMB) radiation. Utilizing numerical compressible MHD calculations and a Monte Carlo Markov chain analysis, which compares calculated CMB anisotropies with those observed by the \textit{WMAP} and \textit{Planck} satellites, we derive limits on the magnitude of putative PMFs. We find that the {\it total} remaining present day field, integrated over all scales, cannot exceed 47 pG for scale-invariant PMFs and 8.9 pG for PMFs with a violet Batchelor spectrum at 95\% confidence level. These limits are more than one order of magnitude more stringent than any prior stated limits on PMFs from the CMB which have not accounted for this effect.
\end{abstract}

\maketitle

The early Universe may well have been magnetized. There is a plethora of proposed magnetogenesis scenarios typically acting well before the epoch of cosmological recombination. These fall into roughly two broad classes: (i) generation of magnetic fields during phase transitions, leading to very blue or violet spectra and (ii) generation of magnetic fields during
inflation leading to approximately scale-invariant spectra (cf.~Ref.~\cite{Durrer:2013pga} for a review). Even though none of these scenarios is more compelling than others, if only one of them leads to a present day void magnetic field of $\sim 0.005$\,nG \cite{footnote}, the origin of cluster magnetic fields of approximately microgauss strength would be explained immediately~\cite{Banerjee:2003xk,Dolag:2002bw,Banerjee:2004df}. This is irrespective of the correlation length of such fields as long as it is on astrophysical scales, i.e.~in kiloparsec to Megaparsec range \cite{Durrer:2013pga}. An alternative for the origin of cluster magnetic fields is the amplification of astrophysical seed magnetic fields by dynamo action. In any case, independent of the origin of cluster magnetic fields, the question of a potential primordial cosmic magnetization is interesting in its own right.

In fact, fairly recent observations of TeV blazars~\cite{Neronov:1900zz,Tavecchio:2010mk,Takahashi:2013lba} may be best understood if an essentially cosmic volume filling magnetic field with an astrophysical correlation length exists. TeV gamma-rays emitted by these blazars are expected to pair produce $e^{\pm}$ on the extragalactic infrared background~\cite{Plaga:1995ins}, with the resulting $e^{\pm}$ subsequently inverse Compton scattering on the cosmic microwave background radiation (CMB hereafter) to produce secondary GeV gamma rays. This well-predicted flux of GeV photons is, however, not observed in at least three TeV blazars \cite{Neronov:1900zz}. A straightforward explanation is that the $e^{\pm}$ pairs were deflected out of the light cone due to magnetic fields, though other more exotic explanations exist~\cite{Broderick:2011av,0004-637X-758-2-102,Schlickeiser:2013eca,SavelievIGM}. It is by far not clear whether galactic outflows could "contaminate" the Universe with magnetic fields in an essentially volume filling way.

Given these questions, it is therefore not surprising that many authors have searched for indirect observational signatures of primordial magnetic fields (PMFs hereafter). Big bang nucleosythesis, unfortunately, can constrain PMFs only to be smaller than $\sim \mu$G by using their contribution to the cosmic expansion rate. With the advent of precise observations of CMB anisotropies via balloon and satellite observations such as those by \textit{WMAP} and \textit{Planck}, a multitude of stringent limits on PMFs present around recombination, have been placed. These are summarized in Table \ref{T:1} which shows the obtained limits on scale-invariant PMFs. The effects considered, one by one, are $\mu$ and $y$ distortion of the Planck spectrum by the dissipation of magnetic energy into the plasma~\cite{Jedamzik:1999bm,Zizzo:2005az,Kunze:2013uja,Saga:2017wwr}, anisotropic cosmic expansion~\cite{Barrow:1997mj}, CMB temperature anisotropies on high multipoles $l$ due to Alfv{\'e}n and slow magnetosonic waves~\cite{Subramanian:1998fn,Subramanian:2002nh,Mack:2001gc,Lewis:2004kg,Kahniashvili:2005xe,Chen:2004nf, Lewis:2004ef,Tashiro:2005hc,Yamazaki:2006bq,Giovannini:2006gz,Kahniashvili:2006hy,Giovannini:2007qn,Yamazaki:2010nf,Paoletti:2010rx,Shaw:2010ea,Kunze:2010ys,Caprini:2011cw,Paoletti:2012bb,Ballardini:2014jta,Ade:2015cva,Sutton:2017jgr,Zucca:2016iur}, CMB temperature anisotropies due to heating of the plasma shortly after recombination and the increased optical depth~\cite{Sethi:2004pe,Kunze:2013uja,Kunze:2014eka,Ganc:2014wia,Chluba:2015lpa,Ade:2015cva}, creation of additional CMB polarization anisotropies due to Faraday rotation, vector or tensor perturbations (i.e.~gravitational waves) by PMFs~\cite{Durrer:1999bk,Seshadri:2000ky,Mack:2001gc,Subramanian:2003sh,Mollerach:2003nq,Lewis:2004kg,Scoccola:2004ke,Kosowsky:2004zh,Kahniashvili:2005xe,Pogosian:2012jd,Kahniashvili:2014dfa,Zucca:2016iur,Pogosian:2018vfr,Ade:2015cao}, and non-Gaussianity of the CMB induced by PMFs either in the bispectrum~\cite{Brown:2005kr,Seshadri:2009sy,Caprini:2009vk,Cai:2010uw,Trivedi:2010gi,Brown:2010jd,Shiraishi:2010yk,Shiraishi:2011dh,Shiraishi:2013wua,Ade:2015cva,Hortua:2015jkz} or the trispectrum~\cite{Trivedi:2011vt,Trivedi:2013wqa}, as well as effects on reionization~\cite{Sethi:2004pe,Seshadri:2005aa,Schleicher:2011jj,Vasiliev:2014vpa,Pandey:2014vga}. Generally such constraints are in the nanogauss regime and therefore still far from the $0.005$\,nG quoted above and even further from the derived lower limits from TeV blazars. An exception to this rule is the limit of $0.05$\,nG~\cite{Trivedi:2013wqa} from the trispectrum, which, however, is model dependent since it relies on the existence of an additional magnetic-field-induced inflationary curvature mode~\cite{Bonvin:2013tba}.

\begin{figure}
\includegraphics[scale=0.4]{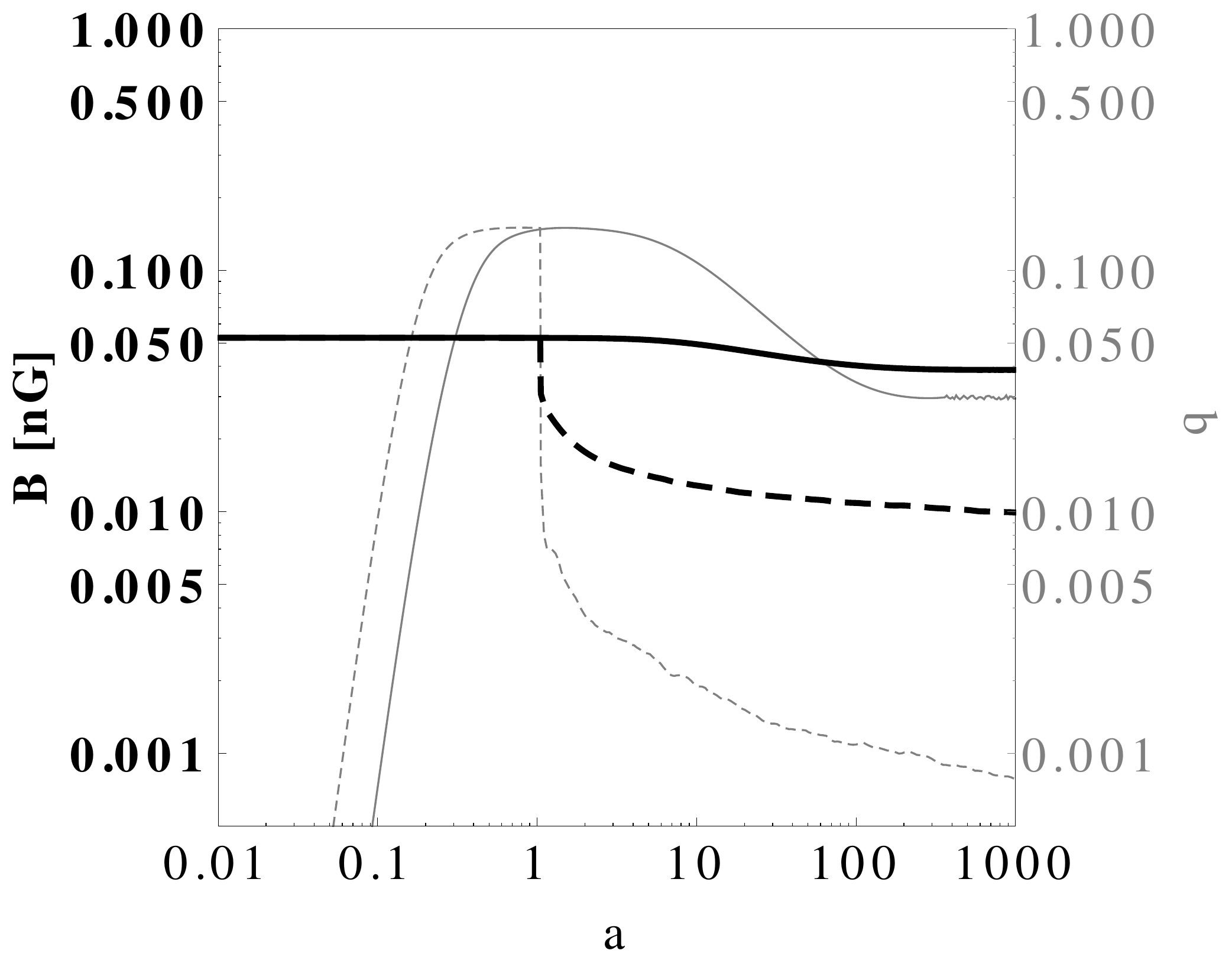}
\caption{The evolution of the magnetic field $B$ (thick black lines, ordinate axis on the left) and the baryon clumping factor $b$ (thin gray lines, ordinate axis on the right) as functions of the scale factor $a$ with $a = 1$ at recombination. The respective solid lines represent the case of a scale invariant ($n = 0$) and the dashed lines that of a Batchelor spectrum ($n = 5$) for the magnetic field.}
\label{fig1}
\end{figure}

\begin{table}[!ht]
\begin{center}
\caption{Constraints on scale-invariant magnetic Fields} 
\begin{ruledtabular}
\begin{tabular}{| c | c |}
Principal effect & Upper limit  \\
& \\ \hline \hline
 Spectral distortions & 30 -- 40 nG \cite{Jedamzik:1999bm,Zizzo:2005az,Kunze:2013uja,Saga:2017wwr} \\ \hline 
 Anisotropic expansion & 3.4nG \cite{Barrow:1997mj}\\ \hline
 CMB temp. anisotropies:& \\ \hline
 Due to magnetic modes & 1.2 -- 6.4 nG \cite{Subramanian:1998fn,Subramanian:2002nh,Mack:2001gc,Lewis:2004kg,Kahniashvili:2005xe,Chen:2004nf, Lewis:2004ef,Tashiro:2005hc,Yamazaki:2006bq,Giovannini:2006gz,Kahniashvili:2006hy,Giovannini:2007qn,Yamazaki:2010nf,Paoletti:2010rx,Shaw:2010ea,Kunze:2010ys,Caprini:2011cw,Paoletti:2012bb,Ballardini:2014jta,Ade:2015cva,Sutton:2017jgr,Zucca:2016iur} \\  
 Due to plasma heating & 0.63 -- 3 nG \cite{Sethi:2004pe,Kunze:2013uja,Kunze:2014eka,Ganc:2014wia,Chluba:2015lpa,Ade:2015cva} \\ \hline 
 CMB polarization & 1.2nG \cite{Durrer:1999bk,Seshadri:2000ky,Mack:2001gc,Subramanian:2003sh,Mollerach:2003nq,Lewis:2004kg,Scoccola:2004ke,Kosowsky:2004zh,Kahniashvili:2005xe,Pogosian:2012jd,Kahniashvili:2014dfa,Zucca:2016iur,Pogosian:2018vfr,Ade:2015cao}\\ \hline  
 Non-Gaussianity bispectrum & 2 -- 9 nG \cite{Brown:2005kr,Seshadri:2009sy,Caprini:2009vk,Cai:2010uw,Trivedi:2010gi,Brown:2010jd,Shiraishi:2010yk,Shiraishi:2011dh,Shiraishi:2013wua,Ade:2015cva,Hortua:2015jkz} \\ \hline
 Non-Gaussianity trispectrum & 0.7nG \cite{Trivedi:2011vt} \\ \hline
 Non-Gaussianity trispectrum &  \\ 
 with inflationary curv.~mode & 0.05nG \cite{Trivedi:2013wqa} \\ \hline  
 Reionization & 0.36 nG \cite{Sethi:2004pe,Seshadri:2005aa,Schleicher:2011jj,Vasiliev:2014vpa,Pandey:2014vga}
\label{T:1}
\end{tabular}
\end{ruledtabular}
\end{center}
\end{table}

The situation is even more bleak for blue and violet spectra, where the magnetic energy resides on small scales; i.e., its correlation length is small. Limits from $\mu$ and $y$ distortions are in the 30\,nG regime~\cite{Jedamzik:1999bm,Zizzo:2005az,Kunze:2013uja,Saga:2017wwr}; however, a limit from the dissipation of magnetic fields after recombination may reach down to values around the subnanogauss regime, though detailed calculations show that it is by far not as stringent
as the limit we will place here \cite{KunzePC}. Placing CMB constraints on PMFs with blue and violet spectra is more difficult than placing them on scale-invariant ones, as small-scale physics is not directly visible in the CMB at multipoles $l\sim 1000$, but rather indirectly. They also rely much more on a solid understanding of the considerable evolution of PMFs on small scales. In particular, the dissipation rates of PMFs before and after recombination have to be known, among others. However, extensive study of decaying magnetohydrodynamics in the expanding Universe in the presence of fluid viscous terms in the linear regime~\cite{Jedamzik:1996wp,Subramanian:1997gi}, by full magnetohydrodynamics (MHD) simulations~\cite{Christensson:2000sp,Banerjee:2004df}, as well as by semianalytic methods~\cite{Jedamzik:2010cy,Saveliev:2012ea,Saveliev:2013uva}, has led to a consistent picture. In the remainder of this letter we will 
establish that PMFs create small-scale baryonic density fluctuations, which
in turn lead to substantial modifications of the CMB anisotropies.

One particular realization coming from PMF evolution studies is that shortly before recombination the part of the spectrum of PMFs that undergoes nontrivial dynamical evolution is well below the photon mean free path at the epoch of recombination. This has the important implication that the effective speed of sound entering the MHD equations is the baryonic and not the radiation one, leading to compressible MHD and the likely creation of $\delta\rho_{\rm b}/\rho_{\rm b}\sim 1$ baryonic inhomogeneities on $\sim$ kiloparsec scales for subnanogauss fields
~\cite{Jedamzik:2011cu,Jedamzik:2013gua}. Imagine a stochastic magnetic field and negligible velocities $\bf v$ initially. The evolution of velocities and densities are given by the Euler and continuity equations
\begin{eqnarray}
\label{eq:fluid1}
\frac{\partial\bf v}{\partial t} + \bigl({\bf v}\cdot {\bf\nabla}\bigr)\cdot{\bf v}
+ c_{\rm s}^{2}\frac{{\bf\nabla} \rho_{\rm b}}{\rho_{\rm b}} & = & - \alpha {\bf v}
-\frac{1}{4\pi\rho_{\rm b}}{\bf B}\times
\bigl({\bf \nabla}\times {\bf B}\bigr)\,,\\
\label{eq:fluid2}
\frac{\partial\rho_{\rm b}}{\partial t} + {\bf \nabla}\bigl(\rho_{\rm b} {\bf v}\bigr) & = & 0 \,, 
\end{eqnarray}
where $\alpha = 4 c \rho_{\gamma}/3\rho_{\rm b} l_{\gamma}$ (cf.~\cite{Banerjee:2004df}) is the photon drag term, with
$\rho_{\gamma}$ and $\rho_{\rm b}$ as the photon and baryonic density, respectively,
$l_{\gamma}$ is the photon mean free path, $c$ is the speed of light, and $c_{\rm s}$ is the baryonic sound speed. Before recombination the fluid is in an overdamped, highly viscous state with the source of viscosity given by free-streaming photons. In this case, only the terms on the rhs of Eq.~(\ref{eq:fluid1}) are important. Very quickly  ($\Delta t\sim 1/\alpha$) terminal velocities of $v\simeq c_{\rm A}^{2}/(\alpha L)$ are reached, with $c_{\rm A} = B/\sqrt{4\pi\rho_{\rm b}}$ being the Alfv{\'e}n velocity of the baryon plasma. For a stochastic field the generated fluid flows are necessarily both rotational (i.e.~$\nabla\times {\bf v}\neq 0$) and compressible (i.e.~$\nabla\cdot{\bf v}\neq 0$). The compressibility component leads to the creation of density fluctuations. Using Eq.~(\ref{eq:fluid2}) one finds $\delta\rho_{\rm b}/\rho_{\rm b} (t)\simeq v t/L\simeq c_{\rm A}^2 t/(\alpha L^2)$. These density fluctuations become larger with time until {\it either} pressure forces become important in counteracting further compression {\it or} the source magnetic stress term decays. The former happens when the last term on the LHS of Eq.~(\ref{eq:fluid1}), $(c_{\rm S}^{2}/L)\,\delta\rho_{\rm b}/\rho_{\rm b}$,  is of the order of the magnetic force term $c_{\rm A}^{2}/L$. That is, density fluctuations may not become larger than $\delta\rho_{\rm b}/\rho_{\rm b} \lsim (c_{\rm A}/c_{\rm s})^2$. The magnetic fields sourcing density fluctuations decay when the eddy turnover rate in the viscous regime $v/L\simeq c_{\rm A}^{2}/\alpha L^{2}$ equals the Hubble rate $H\simeq 1/t$. This has been confirmed by direct numerical simulations~\cite{Banerjee:2004df}, a linear analysis~\cite{Jedamzik:1996wp}, and a particular nonlinear estimate~\cite{Subramanian:1997gi}. Putting all this together, we expect
\begin{equation}
\frac{\delta\rho_{\rm b}}{\rho_{\rm b}}\simeq {\rm min}\,
\biggl[1,\biggl(\frac{c_{\rm A}}{c_{\rm s}}\biggr)^2\biggr]
\end{equation}
for the density fluctuations generated by magnetic fields before recombination. It was further found by analytical estimates based on the results of Ref.~\cite{Banerjee:2004df} that the total magnetic field strength undergoes a drop from $B_{\rm br}$ to $B_{\rm ar}$ (where "br" denotes the value before and "ar" is the value after recombination) of $B_{\rm br}/B_{\rm ar}\approx (\alpha_{\rm rec}/H_{\rm rec})^{n/(2n+4)}$ due to dissipation during and somewhat after recombination, where $\alpha_{\rm rec}/H_{\rm rec}\approx 170$ and $H_{\rm rec}$ is the Hubble constant at recombination. Here $n$ is the spectral index of the PMF, with $n=0$ corresponding to the scale-invariant case.

We now present our results of numerical three-dimensional MHD simulations, i.e. considering {\it compressible} MHD in the early Universe. Note that compressible MHD simulations in the early Universe have been previously performed (see e.g., \cite{Kahniashvili:2012uj,Brandenburg:2014mwa}), however, addressing different physical questions than considered here. The simulations were performed via a novel method of the use of kinetic consistent schemes~\cite{DokMath.90.1.495,RJNAMM.30.1.27,AML.72.75}, which have recently also been successfully applied to astrophysical problems~\cite{CMMPh.55.8.1290,DokMath.95.1.68,DokMath.98.1.396}. Cosmic expansion was included by working with a set of rescaled physical variables~\cite{Banerjee:2004df} (note that the procedure is different than in the case of conformal invariance, cf.~\cite{Brandenburg:1996fc}). Recombination was modeled by a sudden drop in the electron fraction and the concomitant large decrease of $\alpha$. 

A comoving box size of $(10\,{\rm kpc})^{3}$ including an initially homogeneous baryon fluid with zero peculiar velocities and a stochastic, but statistically homogeneous and isotropic magnetic field, was used. It is stressed that such initial conditions should well approximate the physical state of the plasma before recombination
and on the scales considered, i.e.~$L \ll l_{\gamma}$, 
as preexisting adiabatic baryon perturbations have been erased by 
Silk damping earlier on, and peculiar velocity flows, in the absence of other sources, will quickly dissipate due to the strong photon drag, cf.~Eq.~(\ref{eq:fluid2}). Other putative sources of small-scale baryon fluctuations,
such as primordial baryon isocurvature fluctuations, inhomogeneities induced
by cosmic strings, etc., are assumed to be absent, as any further 
inhomogeneities would only strengthen our observational limits found below.

The magnetic field property was described by its Fourier spectrum, i.e.~$\langle |B(k)|^2 \rangle \propto k^{n - 3}$, and its total initial magnetic energy $V B(a_{0})^{2}=\int {\rm d}V B(x,a_{0})^{2}$, with $V$ being the total volume and $a_{0}$ the cosmic scale factor at the beginning of the simulation. For recombination at redshift $z=1090$ we find $c_{\rm s} = 6.33\,$km/s for the isothermal sound speed of fully ionized hydrogen and singly ionized helium with a helium mass fraction $Y_{p}\approx 0.245$, and $c_{\rm A} = 4.34\, {\rm km/s}\,[B_{0}/(0.03\,{\rm nG})]$ for the Alfv{\'e}n velocity. Comparing simulations with $(256)^3$ and $(128)^3$ lattices, we found our results well converged. 

In Fig.~\ref{fig1}, the evolution of the magnetic field strength is shown (thick black lines). Here two initial conditions have been assumed: (i) a violet Batchelor spectrum with $n=5$ and $B(a_{0}) = 52.5\,$pG, where $n=5$ is the strong theoretical expectation for magnetogenesis during phase transitions~\cite{Durrer:2003ja,Saveliev:2012ea}, and (ii) a scale-invariant spectrum of $n=0$ and the same $B(a_{0})$, modeling inflationary produced PMFs. During and after recombination at scale factor $a = 1$, the violet spectral PMF undergoes significant further damping of a factor $5.3$ up to the present epoch. This is in good agreement with the above mentioned theoretical expectation of $6.26$. The scale-invariant field also receives some damping of the total magnetic energy density during and after recombination. However, the exact amount is dependent on the resolution. Though the small-scale dissipative cutoff of the field indeed increases by a factor $\sqrt{\alpha_{\rm rec}/H_{\rm rec}}\approx 13$ across recombination, if fields are excited all the way to Fourier mode $k \to 0$, the energy density would stay essentially the same. This damping factor is therefore not taken into consideration when formulating limits.

The thin gray lines in Fig.~\ref{fig1} show the evolution of the baryonic density fluctuation "clumping factor" $b$, i.e.~$b = (\delta\rho_{\rm b}/\rho_{\rm b}|_{\rm r.m.s.})^2 = (1/V) \int {\rm d}V
\left[\rho_{\rm b}(x)-\langle\rho_{\rm b}\rangle \right]^2/\langle\rho_{\rm b}\rangle^2$. It is seen that in both scenarios (i) and (ii) the initially homogeneous baryon fluid acquires density fluctuations of considerable magnitude before recombination due to magnetic compression. For the assumed $50\,$pG fields this baryon clumpiness exists on the characteristic scale of $c_{\rm A}/(\alpha H)^{1/2}|_{\rm rec}\sim 0.5\,$kpc before recombination. The small-scale baryon inhomogeneity is then very quickly reduced during recombination, though it remains with some lower amplitude up to the present epoch. The decay of inhomogeneities during recombination is due to the almost instantaneous disappearance of the drag $\alpha\ll H$, as electrons recombine into hydrogen, making the fluid enter a fully turbulent MHD evolution.

\begin{figure}
\includegraphics[scale=0.491]{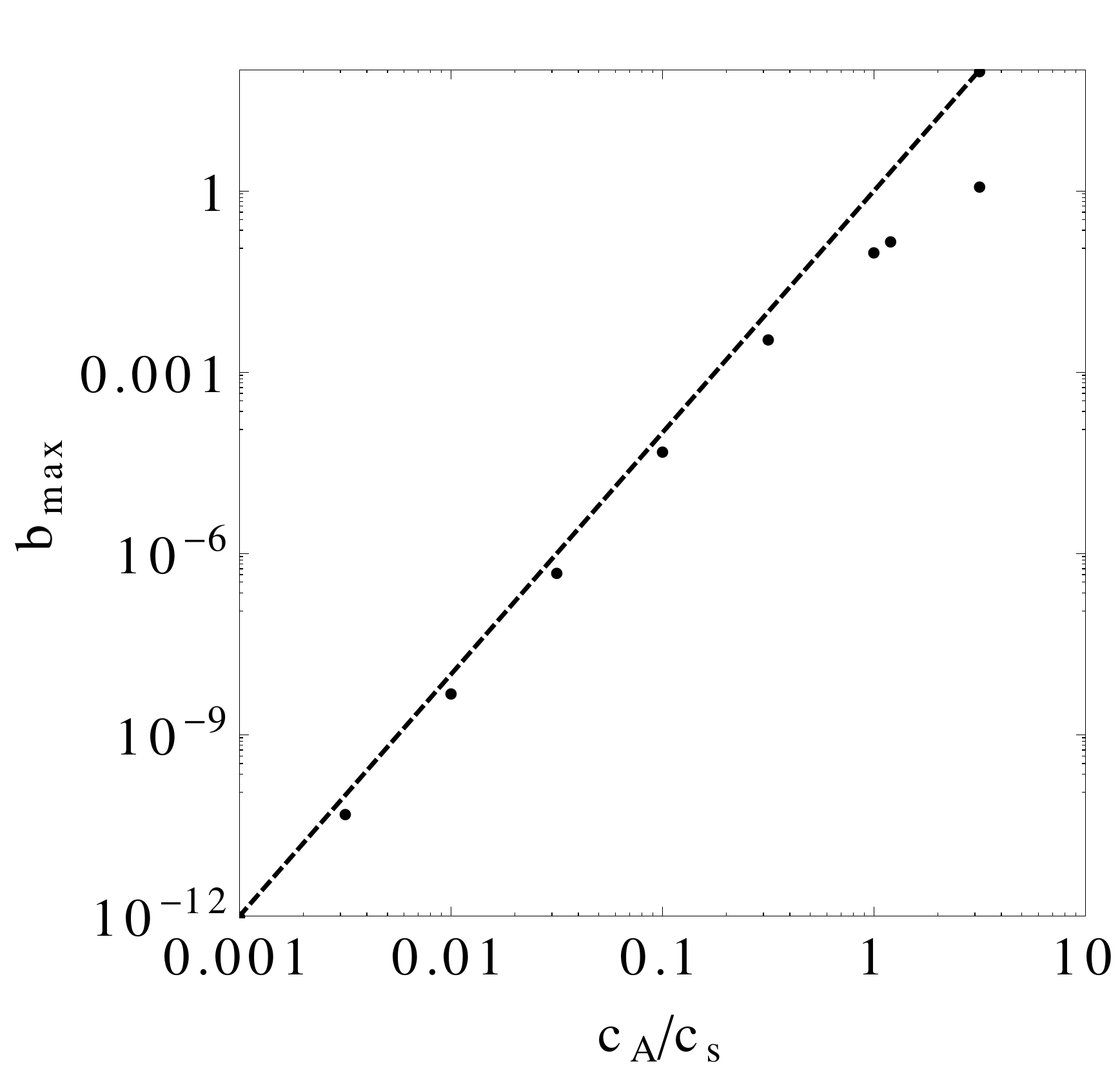}
\caption{The maximum clumping factor $b_{\rm max}$ attained before recombination as a function of $c_{\rm A}/c_{\rm s}$. For reference, the scaling law $b_{\rm max} = (c_{\rm A}/c_{\rm s})^{4}$ is shown as a dashed line.}
\label{fig2}
\end{figure}

From arguments given above it is expected that the maximum clumping before recombination scales as $b\propto (c_{\rm A}/c_{\rm s})^4$ for $c_{\rm A} < c_{\rm s}$ and is constant for $c_{\rm A} > c_{\rm s}$. In Fig.~\ref{fig2} the maximum clumping factor $b_{\rm max}$ is shown as a function of $(c_{\rm A}/c_{\rm s})$ and confirms the fourth power scaling up to ratios of $(c_{\rm A}/c_{\rm s})\sim 0.3$ and a slow turnover for larger ratios.

Small-scale ($L\ll l_{\gamma}$) baryon inhomogeneities
significantly affect the observable CMB anisotropies on large scales $L\gg l_{\gamma}$~\cite{Jedamzik:2011cu,Jedamzik:2013gua}. This is because
the photon mean free path, determined by Thomson scattering of photons on free 
electrons, is changed due to a change of the ionization history. 
The free electron density is determined by a competition between the
recombination rate and the ionization rate. Here the former is proportional
to $\rho_{\rm b}^{2}$, whereas the latter is proportional to $\rho_{\rm b}$ (cf.~\cite{Jedamzik:2011cu,Jedamzik:2013gua}). 
In an inhomogeneous universe we have
$\langle\rho_{\rm b}^{2}\rangle > \langle\rho_{\rm b}\rangle^{2}$, such that average
recombination is stronger, while average ionization stays the same.
This leads to a lower free electron density and a larger photon mean free path,
which in turn leads to enhanced Silk damping and earlier recombination.
It is important to note that the typical scale of the fluctuations is of no relevance, as long as the scale is much below $l_{\gamma}$.
In our analysis we compute the average ionization as a properly
weighted ionization of different regions with different baryon densities. The
statistics of these overdensities is described by the clumping factor $b$. 
This averaged photon mean free path then
enters the computation of the CMB anisotropies, performed with the publicly available code CAMB~\cite{Lewis:1999bs}.

\begin{figure}
\includegraphics[scale=0.491]{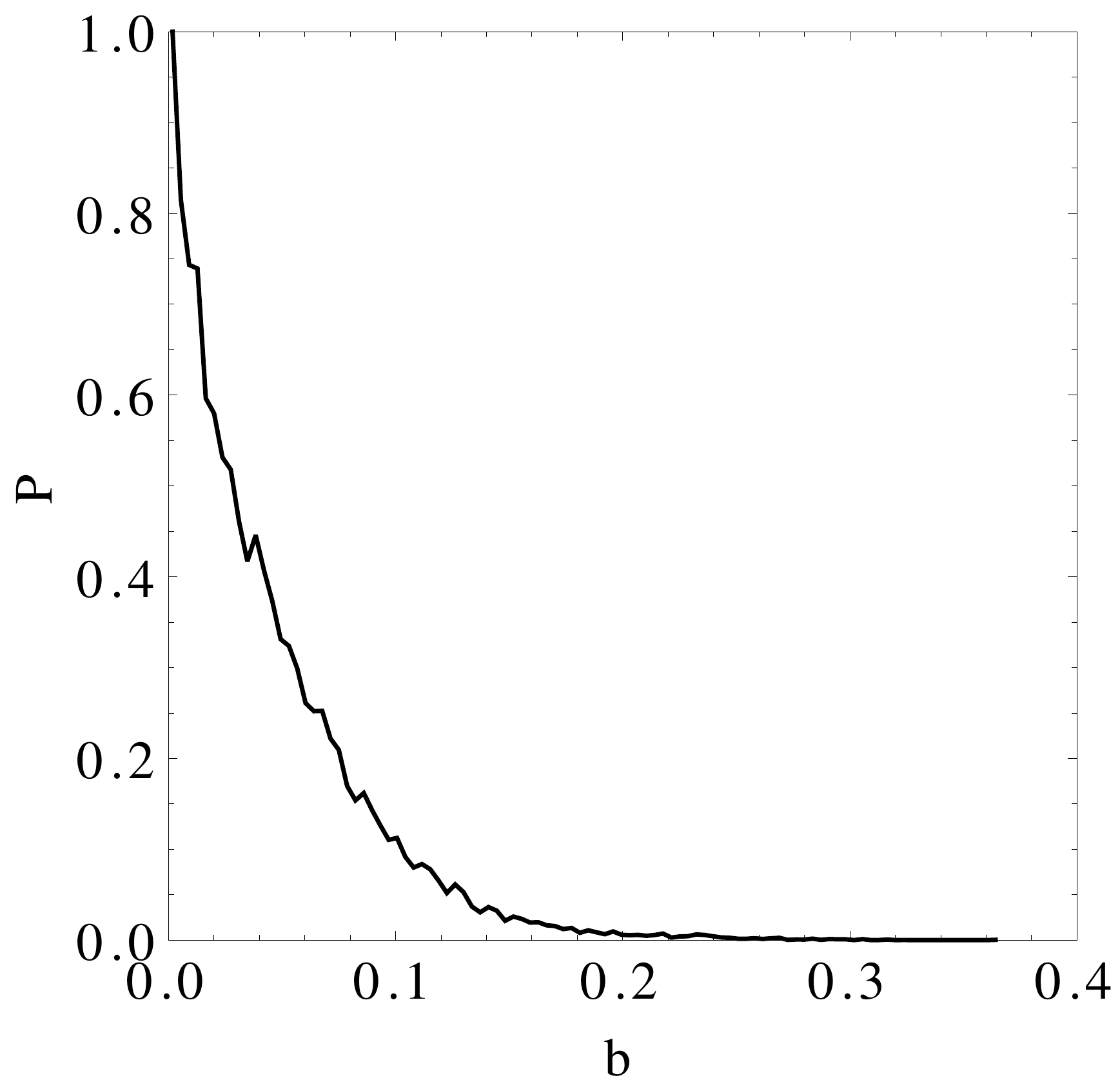}
\caption{The \textit{a posteriori} probability $P$ that a universe with baryon clumps before recombination leads to observationally acceptable CMB temperature anisotropies when compared to \textit{Planck} and \textit{WMAP} data as a function of baryonic clumping factor $b$.}
\label{fig3}
\end{figure}

We have performed an extensive Markov chain Monte Carlo (MCMC) simulation to compare the thus predicted anisotropies of the CMB with the ones observed by Planck \cite{Ade:2013sjv} and WMAP \cite{Bennett:2012zja}, in a cosmic model described by six standard cosmic parameters, but also including small-scale inhomogeneities such as those produced by PMFs. The sole effect of such small-scale inhomogeneities was assumed to be the change in the recombination history. Our analysis was performed by using the CosmoMC generator \cite{Lewis:2002ah}. Figure \ref{fig3} shows the observed \textit{a posteriori} probability for such baryonic clumping with clumping factor $b$ to present a good fit to the data when marginalizing over all six other standard parameters. It is seen that $b$ is limited to $b<0.119$ at 95\% confidence level. Unfortunately, there is no evidence for baryonic clumping, or indirectly for the existence of PMFs, such that $b=0$ gives the best fit.

These results, in conjunction with the results shown in the other figures, may be used to determine a precise limit on PMFs from inhomogeneous recombination. Note that the PMF scenarios that are shown in Fig.~\ref{fig1} produce a maximum clumping of $b = 0.15$ and are therefore excluded somewhat beyond the 95\% confidence level. The 95\% confidence level excluded PMFs are given by
\begin{eqnarray}
\nonumber
B & < & 47\,{\rm pG}\quad\quad {\rm scale-invariant\,\, spectra\,\,} n=0\,, \\
\nonumber
B & < & 8.9\,{\rm pG}\quad\quad {\rm Batchelor\,\, spectrum\,\,} n=5\,.
\end{eqnarray} 
It is stressed here that the quoted limits are on the {\it total} magnetic field, integrated over all scales.

In summary, we have confirmed in detail the suggestion made in Refs.~\cite{Jedamzik:2011cu,Jedamzik:2013gua} that small-scale, comparatively weak primordial magnetic fields may create substantial small-scale baryon density fluctuations which cause the Universe to recombine inhomogeneously. This inhomogeneous recombination in turn may alter the large-scale cosmic microwave background temperature anisotropies to an observable degree. By full numerical compressible MHD simulations, numerical calculations of the resultant CMB anisotropies and Monte Carlo Markov chain analysis of the \textit{Planck} and \textit{WMAP} data we have been able to place the, to date, most stringent limits on the total surviving primordial magnetic field. These limits are about 1--2 orders of magnitude more stringent for inflationary produced fields, and 2--3 orders of magnitude for "causally" produced fields, than a host of other stated CMB constraints on primordial magnetic fields. It is noteworthy that the derived limit for violet spectra is close to the required value for primordial magnetic fields to explain the origin of cluster magnetic fields.  

\begin{acknowledgments}
The work of A.S.~was supported by the Russian Science Foundation under Grant No.~19-11-00032. We are grateful to Kerstin Kunze and Levon Pogosian for valuable discussions.
\end{acknowledgments}


\begin{thebibliography}{10}
\bibitem{footnote}
In this paper, all magnetic field strength and length scales are quoted with their comoving values, i.e.~they would have at the present epoch after cosmic expansion.
\end{thebibliography}
\end{document}